\documentclass{revtex4}
\usepackage{amssymb}
\usepackage{amsmath}
\usepackage{graphicx}
\usepackage[font={footnotesize,it}]{caption}

\begin{document}

\title[ ]{Modified Rindler acceleration as a nonlinear electromagnetic effect%
}
\author{M. Halilsoy}
\email{mustafa.halilsoy@emu.edu.tr}
\author{O. Gurtug}
\email{ozay.gurtug@emu.edu.tr}
\author{S. H. Mazharimousavi }
\email{habib.mazhari@emu.edu.tr}
\affiliation{Department of Physics, Eastern Mediterranean University, G. Magusa, north
Cyprus, Mersin 10, Turkey. }

\begin{abstract}
The model proposed originally by Mannheim and Kazanas for fitting the shapes
of galactic rotation curves has recently been considered by Grumiller to
describe gravity of a central object at large distances. Herein we employ
the same geometry within the context of nonlinear electrodynamics (NED).
Pure electrical NED model is shown to generate the novel Rindler
acceleration term in the metric which explains anomalous behaviors of test
particles / satellites. Remarkably a pure magnetic model of NED yields flat
rotation curves that may account for the missing dark matter. Weak and
Strong Energy conditions are satisfied in such models of NED.
\end{abstract}

\maketitle

\section{INTRODUCTION}

In Newton's theory of gravitation which reined for centuries, mass
constituted the principal source for its potential. With the advent of
general relativity, different sources were unified under spacetime texture
such that the overall effective force used to matter. Thus, gravity /
geometry can easily be attributed to non-mass originated sources equally
well due to the manifestation of mass-energy equivalence. As a particular
example we recall the Reissner-Nordstr\"{o}m (RN) geometry of general
relativity in which mass and charge coexist in making the geometry. Assuming
that the source has negligible mass versus a significant charge the entire
geometry can be attributed to the charge alone. In performing this process
one should be cautious that no physical energy conditions are violated.
Recent observations suggest that there are dark matter / energy that is
associated with non-observable sources. As a result our detectable /
observable matter falls rather short to account for the accelerated
expansion of our universe. \ Before suggesting proposals for new forces /
matter it is more logical to exhaust every kind of physical sources that we
are at least familiar so that we know how to cope with. To explain the
hierarchy of forces at small distances and close the gap of discrepancy
between gravity and other fields , for instance, the idea of higher
dimensions / branes was proposed \cite{1,2}. Although no branes have been
identified so far theoretical explanation such as dilution (weaking) of
gravity among branes in higher dimensions remains consistently intact. As
the gravity is tamed at UV scales by virtue of higher dimensions at the IR
scales, or long distances does everything go perfect?. The recent proposal 
\cite{3,4,5,6} that at large distances there is an additional parameter
known as Rindler acceleration was rather unprecedented and the present paper
is about the source of such a term.

We recall that in the near horizon limit, i.e. $r=2m+x$ for $\left\vert
x\right\vert ^{2}\ll 1,$ a Schwarzschild black hole leads to the standard
Rindler acceleration. Such an extraneous term must be purely general
relativistic coupled with physical sources which lacks a Newtonian
counterpart. Even in the Einstein-Maxwell version of general relativity with
spherical symmetry such a term did not arise. The long range fields, i.e.
gravitation and electromagnetism, manifest their inverse square law
character so that asymptotically the spacetime becomes flat. Different
sources such as dilatons, nonlinear electromagnetic fields and others admit
non asymptotically flat solutions at large spatial distances. The difficulty
with the new Rindler acceleration is that it violates both the Newtonian and
Maxwellian limits: for large distances ($r\rightarrow \infty $) it becomes
even more significant. In Newtonian terms the potential that gives inverse
force law modifies into $\phi \left( r\right) =-\frac{m}{r}+ar$, where $m$
is the central Newtonian mass and $a$ is the novel Rindler acceleration
under question. Unless the central object is supermassive and $a$ is
negligibly small it can be argued that for large $r$ the new term dominates
over the mass term. Further, the Rindler acceleration is not a universal
constant as observationally it shows slight variations from Sun-Pioneer pair
($\sim 10^{-61}$ natural unit of acceleration which is equivalent to $%
10^{-10}\frac{m}{s^{2}}$ in physical units) to galaxy-Sun system ($\sim
10^{-62}$) and others. We recall that such a linear dependence of potential
on distance is encountered in parallel plates endowed with a uniform
electric field in linear Maxwell electromagnetism (i.e. $V_{0}=E_{0}z$, $%
E_{0}=$ constant).

Gravity coupled with linear Maxwell electromagnetism in spherically
symmetric geometry produces no such linear potential term either. For this
reason we resort from the outset to nonlinear electrodynamics (NED) and
prove a theorem to generate the new acceleration term. Truly it yields the
required expression, however, in addition it gives as a by product an extra
constant term in the metric which can be interpreted as a global monopole 
\cite{7,8}. This amounts to further modification of the Newtonian potential
by $\phi \left( r\right) =k-\frac{m}{r}+ar$, with the global monopole term \ 
$k=$ constant. Our formalism suggests that both the Rindler acceleration ($a$%
) and global monopole ($k$) constants depend on the nonlinear electric
charge of the heavenly object under consideration. That is, neither one is a
fundamental constant of nature as both are derived from the charge.
Interestingly the monopole term plays the similar role of a cosmological
constant, i.e. a uniform electric field in the presence of NED-coupled
gravity with nonisotropic difference. The upper bounds for both $\left\vert
a\right\vert $ and $\left\vert k\right\vert $ have been tabulated for
different planets. It is further shown that the monopole term is crucial for
the weak and strong energy conditions to be satisfied. The spectrum of NED
theories is very large and the problem is to find the proper Lagrangian that
suits and serves for the purpose. Finally we observed that the Rindler
acceleration doesn't account for the constant tangential velocity of
circular orbits in the presence of mysterious dark matter. For this reason
we have further modified the Rindler term in the metric function by $%
2ar\rightarrow 2ar_{0}\ln r$ (with $a$ and $r_{0}$ constant), which
necessitates a new NED Lagrangian. For such a magnetic Lagrangian it is
shown that the energy conditions are satisfied at the cost of a bounded
universe. Further, the circular orbit around remote galaxies, has velocity $%
v=\sqrt{\frac{m}{r}+ar_{0}}$ which yields a better estimate between Newton
and Rindler acceleration models toward accounts of dark matter.

\section{The Solutions}

\subsection{Pure Electric case}

Recently, Grumiller considered the Mannheim-Kazanas (MK) metric to describe
gravity of a central mass at large distances which attracted interest due to
its cosmological implications \cite{5,6}. We must add that a linear term in
the metric was first introduced in \cite{3}, and it was applied in earnest
in fitting the shapes of galactic rotation curves by Mannheim (see \cite{4},
for a review). The novelty in this model is the inclusion of a term
interpreted as Rindler acceleration. We wish to show in this paper that
nonlinear electrodynamics (NED) may be responsible for the generation of
such a term. Our starting point is the action 
\begin{equation}
S=\int d^{4}x\sqrt{-g}\left[ R-2\Lambda +L\left( \mathcal{F}\right) \right]
\end{equation}%
in which $R=$ Ricci scalar, $\Lambda =$ cosmological constant and $L\left( 
\mathcal{F}\right) $ is the Lagrangian for the NED.

Before we choose the form of $L\left( \mathcal{F}\right) $ we would like to
add that $L\left( \mathcal{F}\right) $ is not similar to the original BI
Lagrangian. In Born-Infeld (BI) initial work the idea was the removal of the
singularity at the origin. Following the classical charge with a finite size
and a well defined charge distribution admitted what we call it BI
Lagrangian. In what we introduce the singularity at the origin is not our
worry any more and instead we are adjusting our Lagrangian to justify the
behavior of the Galaxies at very large distance. Hence, the only constraint
we impose on our Lagrangian is to satisfy the Maxwell equation with a single
electric or magnetic fields. No need to mention that such an arbitrary
Lagrangian may not give the Maxwell limit at large distance which otherwise
expecting the Mannheim-Kazanas instead of Reissner-Nordstr\"{o}m would be
meaningless.

Our notation is such that $\mathcal{F}=F_{\mu \nu }F^{\mu \nu \text{ }}$
represents the Maxwell invariant with the choice of Lagrangian 
\begin{equation}
L\left( \mathcal{F}\right) =\frac{\alpha }{\sqrt{2}\beta -\sqrt{-\mathcal{F}}%
}.
\end{equation}%
Here $\alpha >0$ is the coupling constant and $\beta >0$ plays the role of
the uniform background electric field as will be clarified in the sequel.
Our original model Lagrangian (2) can be employed with the choice $\beta =0$
as well. Note that $F_{\mu \nu }=\partial _{\mu }A_{\nu }-\partial _{\nu
}A_{\mu }$ is the standard Maxwell field tensor and this Lagrangian will
break the scale invariance, i.e. $x\rightarrow \lambda x,$ $A_{\mu
}\rightarrow \frac{1}{\lambda }A_{\mu },$ for $\lambda =$ constant. We
consider a static, spherically symmetric (SSS) spacetime described by the
line element 
\begin{equation}
ds^{2}=-f(r)dt^{2}+\frac{dr^{2}}{f(r)}+r^{2}\left( d\theta ^{2}+\sin
^{2}\theta d\phi ^{2}\right) ,
\end{equation}%
with the electric field ansatz.

The fact that the desired static spherically symmetric (SSS) line element
(3) derives from the NED Lagrangian (2) through the action (1) can be
formulated as a theorem \cite{9}.

\textbf{Theorem:} Let our action be (1) with line element (3) and $L(F)$ be
our pure electric NED Lagrangian described by the Maxwell $2-$form%
\begin{equation}
\mathbf{F}=E\left( r\right) dt\wedge dr
\end{equation}%
satisfying the Maxwell's equation%
\begin{equation}
d\left( ^{\star }\mathbf{F}L_{\mathcal{F}}\right) =0
\end{equation}%
in which $^{\star }\mathbf{F}$ means dual of $\mathbf{F}$ and $L_{\mathcal{F}%
}=\frac{\partial L}{\partial \mathcal{F}}.$ Then, the energy-momentum tensor
satisfies the conditions $T_{t}^{t}=T_{r}^{r}$ and $T_{\theta }^{\theta
}=T_{\varphi }^{\varphi }$ and Lagrangian $L(\mathcal{F})$ is related to the
metric function $f(r)$ through%
\begin{equation}
L=L_{0}+2\int \frac{1}{r^{2}}\left[ r^{2}\left( \frac{f^{\prime \prime }}{2}+%
\frac{1-f}{r^{2}}\right) \right] ^{\prime }dr
\end{equation}%
where $L_{0}=const.$ and a 'prime' implies $\frac{d}{dr}.$

\textbf{Proof:} Variation of the action (1) with respect to the metric
tensor $g_{\mu \nu }$ yields the field equations in the form%
\begin{equation}
G_{\mu }^{\nu }+\Lambda \delta _{\mu }^{\nu }=T_{\mu }^{\nu }
\end{equation}%
where $G_{\mu }^{\nu }$ is the Einstein tensor and $T_{\mu }^{\nu }$ is the
energy-momentum tensor given by%
\begin{equation}
T_{\mu }^{\nu }=\frac{1}{2}\left( L\delta _{\mu }^{\nu }-4L_{\mathcal{F}%
}F_{\mu \lambda }F^{\nu \lambda }\right)
\end{equation}%
which admits $T_{t}^{t}=T_{r}^{r}=\frac{1}{2}L-\mathcal{F}L_{\mathcal{F}}$
and $T_{\theta }^{\theta }=T_{\varphi }^{\varphi }=\frac{1}{2}L.$ From the
line element (3) we find%
\begin{equation}
G_{t}^{t}=G_{r}^{r}=\frac{rf^{\prime }-1+f}{r^{2}}
\end{equation}%
and 
\begin{equation}
G_{\theta }^{\theta }=G_{\varphi }^{\varphi }=\frac{rf^{\prime \prime
}+2f^{\prime }}{2r}.
\end{equation}%
The electric field $2-$form has the dual given by%
\begin{equation}
^{\star }\mathbf{F}=-E\left( r\right) r^{2}\sin \theta d\theta \wedge
d\varphi
\end{equation}%
and the Maxwell's equation (5) implies that%
\begin{equation}
E\left( r\right) r^{2}L_{\mathcal{F}}=const.=Q
\end{equation}%
in which $Q$ is a charge related integration constant. Recall that, $L_{%
\mathcal{F}}=\frac{\partial L}{\partial \mathcal{F}}=\frac{L_{E}}{\frac{%
\partial \mathcal{F}}{\partial E}}$ and since $\mathcal{F}=-2E^{2},$ we have 
$L_{\mathcal{F}}=-\frac{L_{E}}{4E}.$ Comparing this with Eq. (12) yields%
\begin{equation}
L_{E}=\frac{-4Q}{r^{2}}.
\end{equation}%
From (7) we have $G_{t}^{t}=T_{t}^{t}-\Lambda ,$ which reads%
\begin{equation}
\frac{rf^{\prime }-1+f}{r^{2}}=\frac{1}{2}L-\mathcal{F}L_{\mathcal{F}%
}-\Lambda
\end{equation}%
or alternatively%
\begin{equation}
\frac{rf^{\prime }-1+f}{r^{2}}=\frac{1}{2}\left( L+\frac{4QE}{r^{2}}\right)
-\Lambda .
\end{equation}%
The other field equation $G_{\theta }^{\theta }=T_{\theta }^{\theta
}-\Lambda $ reads as 
\begin{equation}
\frac{rf^{\prime \prime }+2f^{\prime }}{2r}=\frac{1}{2}L-\Lambda .
\end{equation}%
Next, we subtract (16) from (15) which determines the electric field 
\begin{equation}
E=-\frac{r^{2}}{2Q}\left( \frac{f^{\prime \prime }}{2}+\frac{1-f}{r^{2}}%
\right) .
\end{equation}%
Note that from the chain rule we have 
\begin{equation}
\frac{dL}{dr}=-\frac{4Q}{r^{2}}\frac{dE}{dr}.
\end{equation}%
Using (17) and the latter relation one finds%
\begin{equation}
\frac{dL}{dr}=\frac{2}{r^{2}}\left[ r^{2}\left( \frac{f^{\prime \prime }}{2}+%
\frac{1-f}{r^{2}}\right) \right] ^{\prime }
\end{equation}%
or consequently 
\begin{equation}
L=L_{0}+2\int \frac{1}{r^{2}}\left[ r^{2}\left( \frac{f^{\prime \prime }}{2}+%
\frac{1-f}{r^{2}}\right) \right] ^{\prime }dr
\end{equation}%
which completes the proof.

Now we apply the theorem for the MK metric which admits $L\sim \frac{1}{%
\sqrt{-\mathcal{F}}}$. Next, to make our study more general, we modify this
Lagrangian as given in Eq. (2). The nonlinear Maxwell equation $d\left(
^{\star }\mathbf{F}\frac{\partial L}{\partial \mathcal{F}}\right) =0,$
admits the electric field%
\begin{equation}
E(r)=E_{0}-\xi r
\end{equation}%
in which $E_{0}=const.(=\beta )$ and $\xi =2^{-5/4}\sqrt{\frac{\alpha }{C}}%
=const.$ We note that the integration constant $C$ is identified as the
total charge $Q$ of the central object which is obtained from the Gauss's
law $\oint \left( ^{\star }\mathbf{F}\frac{\partial L}{\partial \mathcal{F}}%
\right) =4\pi Q.$ The solution of Einstein-NED equations gives the following
metric function%
\begin{equation}
f(r)=1+2k-\frac{2m}{r}+2ar-\frac{1}{3}\Lambda r^{2},
\end{equation}%
where $k=-QE_{0}=const.<0$, $a=2^{-5/4}\sqrt{\alpha Q}=const.$ while $m=$
mass and $\Lambda =$ the cosmological constant, require no comments. As the
expressions suggest $\alpha $ and $Q$ must have the same sign and the fact
that the acceleration admits both signs has cosmological implications. If
this metric is compared with MK one, it can be easily seen that the
Rindler's acceleration constant is derived from the charge $Q$. Beside this
acceleration we have an additional constant $k=QE_{0}$ which can be
interpreted as a global monopole term \cite{7,8}. The global monopole charge
is identified as $\eta =\pm \sqrt{\left\vert \frac{QE_{0}}{4\pi }\right\vert 
}$ which gives rise to non-radial stresses. Let us add that $Q$ and $E_{0}$\
are both small enough to elude experimental observations. It's origin can be
traced back to big bang as a topological defect. Being coupled to the
distance $\sim r$ from the center of attraction, however, \ enhances its
role at large distances. We recall that the cosmological constant $\Lambda $
is also very small but it couples with $\sim r^{2}$ in the metric to account
for a significant effect. Naturally such a monopole charge modifies the
Newtonian potential by $\Phi \left( r\right) =k-\frac{m}{r},$ which violates
asymptotic flatness. In case that $Q\rightarrow 0$ the line element reduces
to the standard Schwarzschild-de Sitter, as it should be. The essential
parameters of our model consist of $Q,$ $E_{0},$ $m$ and $\Lambda $ (or $k$%
). With the choice $\beta =0$ the uniform electric field $E_{0}$ will not
exist any more. To what extent our model is physical?. Satisfaction of the
energy conditions are vitally important for a physically acceptable
solution. Let us note that in the original MK's model unless $a<0$ the
energy conditions are violated. In the present case with the NED as source
we wish to show that the energy conditions are satisfied. To illustrate this
we explicitly present the corresponding energy momentum tensor as follows%
\begin{equation}
T_{\mu }^{\nu }=\text{diag}\left[ -\rho ,p_{r},p_{\theta },p_{\phi }\right] =%
\frac{2a}{r}\text{diag}\left[ 2+\frac{k}{ar},2+\frac{k}{ar},1,1\right] .
\end{equation}%
The weak energy condition (WEC) requires that $\rho \geq 0,$ and $\rho
+p_{i}\geq 0.$ For the strong energy condition (SEC), \ in addition to WEC,
we must have $\rho +\sum\limits_{i=1}^{3}p_{i}\geq 0.$ These conditions are
both satisfied provided that 
\begin{equation}
2ar\leq \left\vert k\right\vert
\end{equation}%
for our choice of $E_{0}>0$, $Q>0$ so that $k<0$ and $a>0.$ This suggests
that validity of the energy conditions confines the motion by $E_{0}$ and
the Rindler acceleration $a.$ The fact that $a$ is very small ($\simeq
10^{-10}\frac{m}{s^{2}}$) makes $r$ from (24) still quite large. $E_{0}$ can
be interpreted as a uniform background electric field filling all space
which appears also in the Lagrangian (i.e. $E_{0}=\beta $). This makes $%
E_{0} $ an indispensable parameter of the theory provided we stipulate the
energy conditions. Once we set $E_{0}=\beta =0,$ the nonlinear Lagrangian
reduces to $L\left( \mathcal{F}\right) =\frac{-\alpha }{\sqrt{-\mathcal{F}}}$
with the solution for the electric field $E(r)=-\xi r,$ ($\xi =const.$).
However, this will violate the energy conditions and due to this fact, we
are compelled to invoke a space filling uniform electric field $E_{0}$ as
regulator, much like the concept of cosmological constant $\Lambda .$ All
galaxies must be considered immersed in such a background $E_{0}$ to
regulate energy conditions. Although locally $E_{0}$ is too small to be
detected globally it effects the geodesics. (This can best be seen from the
above definitions where $\left\vert E_{0}\right\vert =\frac{\left\vert
k\alpha \right\vert }{a^{2}},$ which can be made arbitrarily small with the
weaker coupling parameter $\alpha $).

The role of $E_{0}$ becomes even more transparent if we assume the central
object to host a black hole. The horizon radius shows a steep rise versus $%
E_{0}$ and the Hawking temperature $T_{H}$ which depends also on the Rindler
acceleration reaches saturation for increasing $E_{0}$. That is, no matter
how $E_{0}$ rises, $T_{H}$ reaches a constant value above zero. From the
thermodynamical point of view, $k=\frac{1}{2}$ acts as a point of phase
transition. From physical standpoints the global monopole parameter, $%
\left\vert k\right\vert =$ $QE_{0}$ may be chosen small enough, apt for
perturbative treatment. This is not imperative however, since relaxation of
this condition will naturally yield from the geodesics equation open,
hyperbolic orbits admissible as well. Global monopoles are known to arise
also in modified $f(R)$ theories \cite{7}.

The equation of motion for a charged test particle is given by%
\begin{equation}
\dot{r}^{2}+V_{eff}(r)=\mathcal{E}^{2}
\end{equation}%
in which the 'dot' stands for derivative with respect to proper time and the
effective potential reads%
\begin{equation}
V_{eff}(r)=q_{0}r\left( E_{0}+\frac{\xi }{2}r\right) \left[ 2\mathcal{E-}%
q_{0}r\left( E_{0}+\frac{\xi }{2}r\right) \right] +f(r)\left( 1+\frac{\ell
^{2}}{r^{2}}\right) .
\end{equation}%
Here $\mathcal{E}$ ($=$ energy) and $\ell $ ($=$ angular momentum) are the
constants of motion while $q_{0}$ is the charge of the test particle. The
simplest way to handle this potential analytically is to consider a neutral (%
$q_{0}=0$) particle at larger $r$ with $\Lambda =0.$ The geodesics equation
simplifies to $\dot{r}^{2}+\left( 1+k+ar\right) \simeq \mathcal{E}^{2},$
which integrates to 
\begin{equation}
r(\tau )=\frac{\mathcal{E}^{2}-k-1}{a}-\frac{3}{2}a^{-1/3}\left( \tau -\tau
_{0}\right) ^{2/3}.
\end{equation}%
It is observed that in this model there is not only a maximum radius (i.e.
Eq. (24)) but also a maximum proper time determined by $\mathcal{E}^{2}/a$
and $\left\vert k\right\vert /a.$

The null geodesics equation for $u(\phi )=\frac{1}{r}$ reads \cite{10}, for $%
\Lambda =0$%
\begin{equation}
\frac{d^{2}u}{d\phi ^{2}}+\left( 1-2\left\vert k\right\vert \right)
u=3mu^{2}-a
\end{equation}%
which upon scalings $\phi \rightarrow \sqrt{1-2\left\vert k\right\vert }\phi 
$, $m\rightarrow \frac{m}{1-2\left\vert k\right\vert }$ and $a\rightarrow 
\frac{a}{1-2\left\vert k\right\vert }$ the perturbative solution can be
found. Employing $u=\frac{\sin \phi }{R},$ as the flat space solution with $%
R=$ the minimum light distance, the total bending angle for the photon orbit 
\cite{10} modifies into%
\begin{equation}
2\psi _{0}\approx \frac{4m}{R}\left[ 1+2ma+2\left\vert k\right\vert \left(
1+6ma\right) \right] .
\end{equation}

\subsubsection{Solar system upper bound for $\left\vert k\right\vert $}

In this subsection we apply a similar calculation as in \cite{6} to find an
upper bound for the parameter $\left\vert k\right\vert $ and therefore for
the background electric field. The effective potential for a particle in a
stable circular motion is given by%
\begin{equation}
V_{eff}=\frac{1}{2}\left[ 1+2k-\frac{2M}{r}+2ar\right] \left( 1+\frac{\ell
^{2}}{r^{2}}\right)
\end{equation}%
where at $r=r_{c}$ we have%
\begin{equation}
\frac{dV_{eff}}{dr}=0
\end{equation}%
which implies%
\begin{equation}
\ell ^{2}=\frac{\left( M+ar_{c}^{2}\right) r_{c}^{2}}{%
-3M+ar_{c}^{2}+r_{c}+2kr_{c}}.
\end{equation}%
A small perturbation applied to the stable orbit of the particle will cause
an oscillatory motion with the frequency%
\begin{equation}
\omega _{r}=\left( \frac{d^{2}V_{eff}}{dr^{2}}\right) _{r_{c}}.
\end{equation}%
Since we are interested in the perihelion oscillation frequency it is given
by%
\begin{equation}
\omega _{p}=\frac{\ell ^{2}}{r_{c}^{2}}-\omega _{r}
\end{equation}%
which up to the first order approximation it amounts to%
\begin{equation}
\omega _{p}\simeq \frac{3M^{3/2}}{r_{c}^{5/2}}\left( 1-\frac{r_{c}^{3}}{%
3M^{2}}a-k\left( \frac{13}{2}+\frac{r_{c}}{3M}\right) \right)
\end{equation}%
or in analogy with \cite{5,6,11} 
\begin{equation}
\omega _{p}\simeq \frac{3M^{3/2}}{A^{5/2}\left( 1-e^{2}\right) ^{5/4}}\left(
1-\frac{A^{3}\left( 1-e^{2}\right) ^{3/2}}{3M^{2}}a-k\left( \frac{13}{2}+%
\frac{A\sqrt{1-e^{2}}}{3M}\right) \right) .
\end{equation}%
Herein $A$ stands for the semimajor axis of the ellipse and $e$ is its
eccentricity. As it has been considered in \cite{5,6} the first term is just
the perihelion frequency in general relativity (leading term). This means
that the second and third terms represent the shift of the general
relativity up to first order in $a$ and $k,$ respectively, i.e., 
\begin{equation}
\frac{\Delta \omega _{p}}{\omega _{p}}=\frac{A^{3}\left( 1-e^{2}\right)
^{3/2}}{3M^{2}}a+k\left( \frac{13}{2}+\frac{A\sqrt{1-e^{2}}}{3M}\right) .
\end{equation}%
This in turn suggests that $\left\vert a\right\vert $ is bounded by%
\begin{equation}
\left\vert a\right\vert <\frac{\Delta \omega _{p}}{\omega _{p}}\frac{3M^{2}}{%
A^{3}\left( 1-e^{2}\right) ^{3/2}}
\end{equation}%
and%
\begin{equation}
\left\vert k\right\vert <\frac{\Delta \omega _{p}}{\omega _{p}}\frac{1}{%
\frac{13}{2}+\frac{A\sqrt{1-e^{2}}}{3M}}.
\end{equation}%
Tab. I shows a list of upper bounds for $\left\vert a\right\vert $ and $%
\left\vert k\right\vert $ for different planets which is consistent with 
\cite{5,6}

\begin{table}[th]
\caption{Upper bound for $\left\vert k\right\vert $ and $\left\vert
a\right\vert .$ We note that the first and third rows are in natural units
i.e., $c=G=\hbar =1.$}
\label{table:nonlin}
\centering
\begin{tabular}{ccccccccc}
\hline\hline
Planet & Mercury & Venus & Earth & Mars & Jupiter & Saturn & Uranus & Icarus
\\[1ex] \hline
$\left\vert a\right\vert {\small <}$ & ${\small 2.82\times 10}^{-64}$ & $%
{\small 1.31\times 10}^{-65}$ & ${\small 1.65\times 10}^{-66}$ & ${\small %
0.25\times 10}^{-66}$ & ${\small 0.94\times 10}^{-64}$ & ${\small 0.11\times
10}^{-65}$ & ${\small 0.42\times 10}^{-65}$ & ${\small 0.90\times 10}^{-62}$
\\ 
$\left\vert a\right\vert \left( \frac{m}{s^{2}}\right) {\small <}$ & $%
{\small 0.16\times 10}^{-12}$ & $0.{\small 73\times 10}^{-13}$ & $0.{\small %
92\times 10}^{-14}$ & ${\small 0.14\times 10}^{-14}$ & ${\small 0.52\times 10%
}^{-12}$ & ${\small 0.61\times 10}^{-14}$ & ${\small 0.23\times 10}^{-13}$ & 
${\small 0.50\times 10}^{-10}$ \\ 
$\left\vert k\right\vert {\small <}$ & ${\small 0.48\times 10}^{-11}$ & $0.%
{\small 71\times 10}^{-11}$ & $0.{\small 15\times 10}^{-11}$ & ${\small %
0.53\times 10}^{-12}$ & ${\small 0.26\times 10}^{-8}$ & ${\small 0.99\times
10}^{-10}$ & ${\small 0.15\times 10}^{-8}$ & ${\small 0.36\times 10}^{-8}$ \\%
[1ex] \hline
\end{tabular}
\end{table}

To complete our calculation we used the data provided in \cite%
{12,13,14,15,16} which are the updated version of those given in \cite%
{16,17,18,19} (Tab. II is the summary of these data).

\begin{table}[th]
\caption{Perihelion precessions and the uncertainties taken from 
\protect\cite{12,13,14,15,20}.}
\centering
\begin{tabular}{ccccccccc}
\hline\hline
Planet & Mercury & Venus & Earth & Mars & Jupiter & Saturn & Uranus & Icarus
\\[1ex] \hline
$\delta \Delta \phi (^{\prime \prime }/cy)\sim \Delta \omega _{p}$ & $%
{\small 0.0030}$ & ${\small 0.0016}$ & ${\small 0.00019}$ & ${\small 0.000037%
}$ & ${\small 0.0283}$ & ${\small 0.00047}$ & ${\small 3.90}$ & ${\small 0.8}
$ \\ 
$\Delta \phi (^{\prime \prime }/cy)\sim \omega _{p}$ & ${\small 42.982}$ & $%
8.646$ & $3.84019$ & ${\small 1.35002}$ & ${\small 0.0587}$ & ${\small %
0.01432}$ & ${\small 3.89}$ & ${\small 9.8}$ \\[1ex] \hline
\end{tabular}
\end{table}
Concerning Tab. I it is seen that the smallest bound is given by Mars which
is $\left\vert k\right\vert <0.{\small 53\times 10}^{-12}.$ It is remarkable
to observe that this value of the lower bound for $\left\vert k\right\vert $
is much smaller than its typical value in grand unified theory $10^{-5}$ 
\cite{9}. Also we add that $\left\vert k\right\vert $ is unit-less unlike $a$
which is in $m/s^{2}.$ In Refs. \cite{20,21,22}, additional perihelion
precessions such as the Lense-Thirring effect $\dot{\varpi}_{LT}$ \cite%
{23,24}, the supplementary rates $\Delta \dot{\varpi}$ and the second
Post-Newtonian (2PN) perihelion precessions are discussed. One may
disentangle these various contributions from the uncertainty given in Tab.
II to find a finely tuned upper bound for the parameters $a$ and $k.$ For
instance here we quote Tab. III from Ref. \cite{21} together with our $%
\left\vert k\right\vert .$ 
\begin{table}[th]
\caption{Additional perihelia precessions such as the Lense-Thirring effect $%
\dot{\protect\varpi}_{LT}$, the supplementary rates $\Delta \dot{\protect%
\varpi}$ and 2PN perihelion precessions quoted from \protect\cite{21}. }
\centering
\begin{tabular}{ccccccc}
\hline\hline
Planet & Mercury & Venus & Earth & Mars & Jupiter & Saturn \\[1ex] \hline
$\Delta \dot{\varpi}\left( mas/cy\right) $\cite{13,25} & ${\small -2.0\pm 3.0%
}$ & ${\small 2.6\pm 1.6}$ & ${\small 0.19\pm 0.19}$ & ${\small -0.020\pm
0.037}$ & ${\small 58.7\pm 28.3}$ & ${\small -0.32\pm 0.47}$ \\ 
$\dot{\varpi}_{LT}\left( mas/cy\right) $\cite{23} & ${\small -2.0}$ & $-0.2$
& $-0.09$ & ${\small -0.027}$ & ${\small -7\times 10}^{-4}$ & $-1\times
10^{-4}$ \\ 
$\dot{\varpi}_{2PN}\left( mas/cy\right) $\cite{26,27} & ${\small 7\times 10}%
^{-3}$ & $6\times 10^{-4}$ & $2\times 10^{-4}$ & ${\small 6\times 10}^{-5}$
& $9\times 10^{-7}$ & $9\times 10^{-8}$ \\ 
$\left\vert k\right\vert \left( \text{Natural units}\right) <$ & ${\small %
0.48\times 10}^{-11}$ & $0.71\times 10^{-11}$ & $0.15\times 10^{-11}$ & $%
0.53\times 10^{-12}$ & $0.26\times 10^{-8}$ & $0.99\times 10^{-10}$ \\%
[1ex] \hline
\end{tabular}
\end{table}
After the corresponding subtraction or addition to what we found is given in
Tab. III which is the same as given in Tab. I. The reason can be due to
small corrections which has no significant effect in our approximation.
Therefore using the updated data given in \cite{12,13,14,15} would be
satisfactory to have a general idea of the upper bound for our parameters.

We note that the effects of a Pioneer-type / Rindler acceleration on the
outer parts of the Solar System have been studied in \cite{28,29,30,31}.
Also, the effect of a Pioneer-type acceleration on the perihelion of a
planet was computed for the first time in \cite{32,33,34,35}.

\subsection{Pure magnetic solution and modified MK metric}

In this section we consider not exactly the MK metric but a modified version
of it in which instead of a linear term we have a logarithmic term. As we
shall show in the sequel, such a model admits a velocity-distance curve
which may be closer to the observational data. In addition to that, we
consider a pure magnetic field to be responsible for such logarithmic term
via a non-linear electrodynamic Lagrangian which is not in the standard BI
Lagrangian form but in a form which accommodate such extra term in the
metric function. As we have mentioned in our previous section, our interest
region is not close to the origin but far distance from the center of
galaxies. In order to establish a pure magnetic solution of NED we start
with the following NED Lagrangian 
\begin{equation}
L\left( \mathcal{F}\right) =\alpha \sqrt{\mathcal{F}}\left( 1-\frac{\ln
\left( \frac{\mathcal{F}}{2}\right) }{4}\right)
\end{equation}%
where $\alpha >0$ is the coupling constant and $\mathcal{F}$ is the Maxwell
invariant. Note that these constants are distinct from the ones given in
(2). The vector potential is $A_{\mu }=\delta _{\mu }^{\varphi }P\cos \theta 
$ where $P>0$ is the magnetic charge so that 
\begin{equation}
\mathcal{F}=F_{\mu \nu }F^{\mu \nu }=2\frac{P^{2}}{r^{4}}>0.
\end{equation}%
The energy momentum tensor's components are given by (8) which explicitly
becomes%
\begin{equation}
T_{t}^{t}=T_{r}^{r}=\frac{\alpha \sqrt{2}P}{2r^{2}}\left( 1-\ln \left( \frac{%
\sqrt{P}}{r}\right) \right)
\end{equation}%
and%
\begin{equation}
T_{\theta }^{\theta }=T_{\phi }^{\phi }=\frac{\alpha \sqrt{2}P}{2r^{2}}.
\end{equation}%
Choosing SSS line element (3), the $tt$ component of the Einstein-NED
equations with $\Lambda =0$ yields a solution of the form%
\begin{equation}
f(r)=1-\frac{2m}{r}+\alpha \frac{\sqrt{2}P}{2}\ln \left( \frac{r}{\sqrt{P}}%
\right)
\end{equation}%
in which $m$ is an integration constant. Next, we introduce two new
parameters as $r_{0}=\sqrt{P}$ and $a=\frac{\alpha \sqrt{P}}{2\sqrt{2}}$ and
upon that the metric function is written as%
\begin{equation}
f(r)=1-\frac{2m}{r}+2ar_{0}\ln \left( \frac{r}{r_{0}}\right) .
\end{equation}%
We would like to comment that, expressing the metric function in latter form
enables us to compare it to the standard form of the MK metric which in turn
implies that $a$ is a similar parameter as Rindler acceleration.

It can be checked that the WEC and SEC are satisfied provided $r<\frac{r_{0}%
}{\sqrt{e}}.$ Naturally, in this model the galaxies can't run out infinity.
The Newtonian potential in this model modifies as 
\begin{equation}
\Phi =-\frac{m}{r}+ar_{0}\ln \left( \frac{r}{r_{0}}\right)
\end{equation}%
so that the attractive force $\overrightarrow{F}=-\overrightarrow{\nabla }%
\Phi $ takes the form 
\begin{equation}
\left\vert \overrightarrow{F}\right\vert =\frac{m}{r^{2}}+\frac{ar_{0}}{r}.
\end{equation}%
It worths to mention that a logarithmic potential yielding a $\frac{1}{r}$
extra-force, has been considered in \cite{36,37} in which it was applied to
the Solar System. For circular orbits since $\left\vert \overrightarrow{F}%
\right\vert =\frac{v^{2}}{r}$ for a unit mass particle, this yields the
velocity function as 
\begin{equation}
v=\sqrt{\frac{m}{r}+ar_{0}}.
\end{equation}%
This differs from the Newtonian model ($v=\sqrt{\frac{m}{r}}$) and the model
proposed by Grumiller ($v=\sqrt{\frac{m}{r}+ar}$) \cite{5}. Evidently, for $%
r\rightarrow \infty $ our model has the advantage since $v\rightarrow cons.,$
which is the case believed to be in the presence of dark matter. Fig. 1
displays the three cases openly with the chosen mass and density functions
as described below shortly. No doubt the gap between our model and Newtonian
one corresponds to the invisible dark matter.

\begin{figure}[tbp]
\includegraphics[width=90mm,scale=0.7]{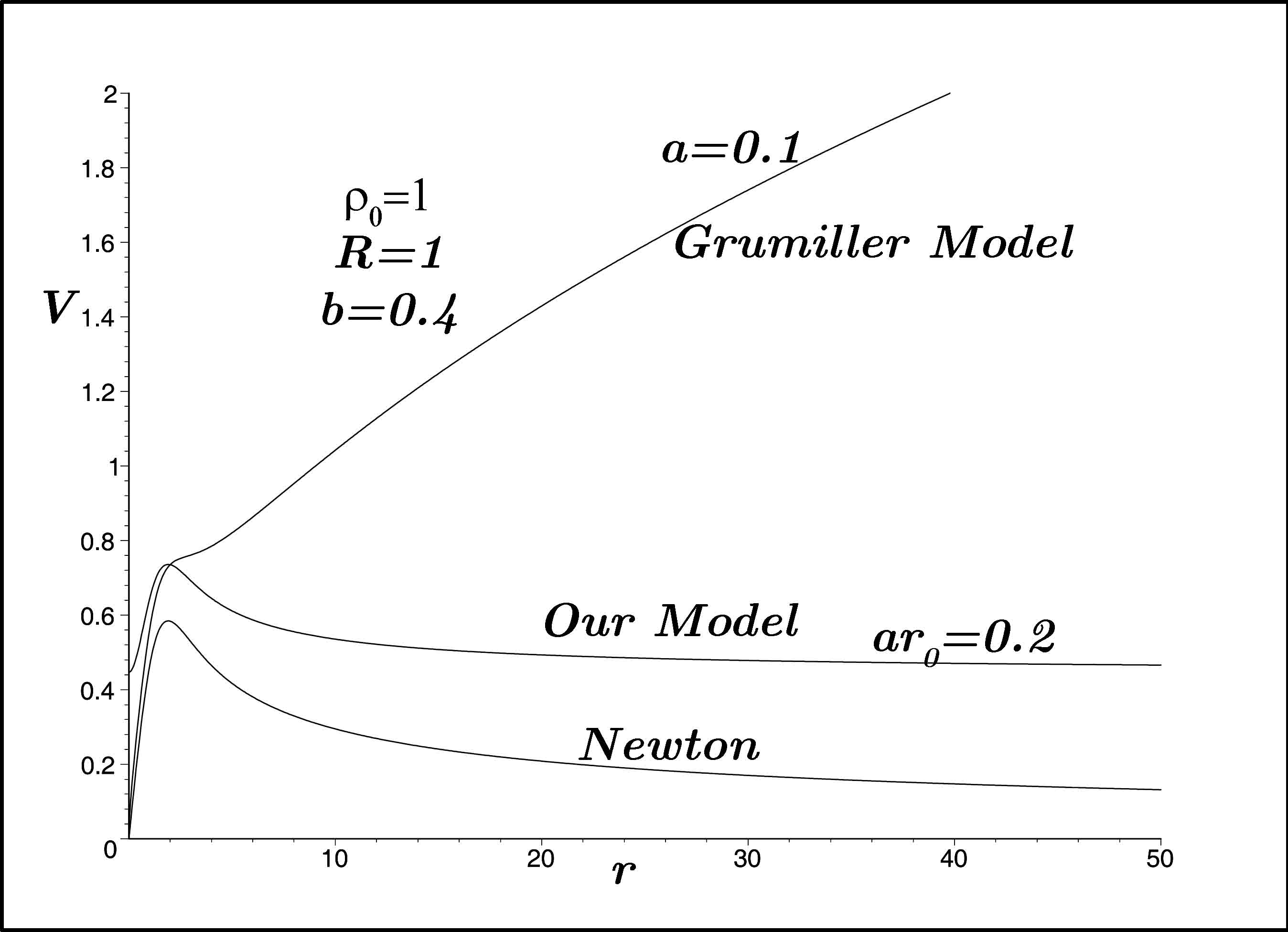}
\caption{A plot of rotational velocity $v(r)$ versus the radial distance for
three different models.}
\end{figure}

Let's consider a model for the mass distribution of a galaxy given by%
\begin{equation}
\rho =\frac{\rho _{0}}{1+\exp \left( \frac{r-R}{b}\right) }
\end{equation}%
in which $\rho _{0},$ $R$ and $b$ are real positive constants. The Fig. 1 is
a plot of rotational velocity $v(r)$ for three different models.

As a side remark, the problem can be considered from the Newtonian
viewpoint. From the Newtonian potential (46) the centripetal force for an
object in circular orbit of radius $r$ is given by%
\begin{equation}
F_{c}=\frac{m}{r^{2}}+\frac{ar_{0}}{r}
\end{equation}%
or equivalently%
\begin{equation}
F_{c}=\frac{m_{n}}{r^{2}}+\frac{m_{d}}{r^{2}}
\end{equation}%
in which $m_{n}=m=$ the normal mass and $m_{d}=ar_{0}r=$ the dark mass. We
assume a normal matter density $\rho _{n}$ and a dark matter density $\rho
_{d}$, such that 
\begin{equation}
m_{n}=4\pi \int\nolimits_{0}^{r}\rho _{n}r^{2}dr\text{ and }m_{d}=4\pi
\int\nolimits_{0}^{r}\rho _{d}r^{2}dr.
\end{equation}%
One easily finds $\rho _{d}=\frac{ar_{0}}{4\pi r^{2}}$ while $\rho _{n}$ is
given by (49). Now, our aim is to see how the parameters should be adjusted
in order to find $\Omega =\frac{m_{d}}{m_{n}}=\frac{23}{4.6}$, i.e. the
experimentally recorded ratio. The parameter $\Omega $ is defined by 
\begin{equation}
\Omega =\left( \int\nolimits_{0}^{\frac{r_{0}}{\sqrt{e}}}\rho
_{d}r^{2}dr\right) /\left( \int\nolimits_{0}^{\frac{r_{0}}{\sqrt{e}}}\rho
_{n}r^{2}dr\right)
\end{equation}%
in which it is assumed that beyond $r_{0}$ both matter and dark matter
become insignificant. The latter equation yields%
\begin{equation}
\int\nolimits_{0}^{\frac{r_{0}}{\sqrt{e}}}\frac{\rho _{0}}{1+\exp \left( 
\frac{r-R}{b}\right) }r^{2}dr=\frac{ar_{0}^{2}}{4\pi \sqrt{e}\eta }.
\end{equation}%
In the zeroth order approximation, one considers a solid central object with
a certain boundary at $r=R$ and a uniform mass distribution $\rho _{0}$
which implies ($b\rightarrow 0$)%
\begin{equation}
\rho _{0}\left( \frac{R^{3}}{3}\right) =\frac{ar_{0}^{2}}{4\pi \sqrt{e}\xi }.
\end{equation}%
Herein $a=\frac{\alpha \sqrt{P}}{2\sqrt{2}},$ $r_{0}=\sqrt{P}$ and upon
substitution it yields%
\begin{equation}
\rho _{0}\left( \frac{R^{3}}{3}\right) =\frac{\alpha P\sqrt{P}}{8\sqrt{2e}%
\pi \Omega }
\end{equation}%
which gives the relation between the radius of the central object $R,$ the
magnetic charge $P$ and coupling constant $\alpha .$ The normal matter $%
m_{n} $ will be expressed in terms of charge $\left( P\right) $ and ratio $%
\Omega $ ($=\frac{23}{4.6}$) by 
\begin{equation}
m_{n}=\frac{\alpha P\sqrt{P}}{2\sqrt{2e}\Omega }.
\end{equation}%
Finally we comment that the effect of dark matter on perihelion precessions
has been also considered in \cite{38}.

\section{CONCLUSION}

In conclusion, the idea of nonlinear electrodynamics (NED) popularized in
1930's by Born and Infeld \cite{39} to resolve singularities remains still
attractive and find rooms of applications even in modern cosmology.
Specifically, a pure electrical NED model serves to generate Rindler
acceleration which was considered responsible for the effects of large
distance gravity. A Theorem has been proved to relate the Lagrangian of NED
with the metric function. Another (i.e. pure magnetic) NED model modifies
the Rindler acceleration term from $\sim 2ar$ to $\sim 2ar_{0}\ln r,$ which
yields better flat rotation curves to conform observations. For a detailed
analysis of geodesics in the presence of the Rindler term we refer to \cite%
{40}. As shown (see Fig. 1), our curve lies in between Grumiller (or MK) and
Newton models. For this reason without resorting to yet unknown particles
dark matter may emerge as a manifestation of NED. The models of NED we
employ here have no counterpart in linear, more familiar Maxwell theory. Our
models are derived in particular to satisfy the energy (Weak and Strong)
conditions and explain the flat rotation curves. As a pay-off in the pure
electric case, for instance a global monopole field crops up which lies
beyond observation for planets in our solar system. This may be considered
much like the cosmological constant, as a background, space-filling uniform
electric field to act as the background energy level. Naturally such fields
are attributed to topological defects as remnants of big bang which are weak
enough to be detected locally. Unfortunately once this field is deleted our
energy conditions will be violated. Finally, it will not be wrong to state
that NED, which has rarely been appealing may encompass larger scopes in
physics / cosmology than envisioned.

\subsection{Acknowledgements}

We wish to thank the anonymous referee for helpful suggestions. A fruitful correspondence with Prof. D. Grumiller is also appreciated.

\end{document}